**Party Comrades and Constituency Buddies: Determinants of Private Initiative Cosponsor Networks in a Parliamentary Multiparty System**

AUTHORS:

Antti Pajala (**corresponding author**)

Doc. Pol. Sci., Adjunct Prof. / Department of Political Science, University of Turku, FI-20014 Turun yliopisto, Finland. Tel. +35823335955 (office), Email: anpaja@utu.fi

Elena Puccio

PhD Student / Dipartimento di Fisica e Chimica, Universita di Palermo, Viale delle Scienze, 90128 Palermo, Italy. Email: elena.puccio@unipa.it

Jyrki Piilo

PhD, Adjunct prof. / Turku Centre for Quantum Physics, Department of Physics and Astronomy, University of Turku, FI-20014 Turun yliopisto, Finland. Email: jyrki.piilo@utu.fi

Michele Tumminello

PhD, Assistant Prof. / Dipartimento di Scienze Economiche, Aziendali e Statistiche, Universita di Palermo, Viale delle Scienze, 90128 Palermo, Italy. Email: michele.tumminello@unipa.it

Acknowledgments: The authors wish to thank Tomi Kause, Timo Forstén and Sakari Nieminen with respect to the data collection.

# Party Comrades and Constituency Buddies; Determinants of Private Initiative Cosponsor Networks in a Parliamentary Multiparty System

**Abstract**

We study Members of Parliament (MP) private initiative (bill) cosponsor patterns from a European parliamentary multiparty perspective. By applying network detection algorithms, we set out to find the determinants of the cosponsorship patterns. The algorithms detect the initiative networks core communities, after which the variables characterizing the core communities can be analyzed. We found legislative network communities being best characterized by the MPs' party affiliations. The budget motion networks, which constitute roughly half of the data, were found mostly characterized by the MPs' home constituencies and only to a limited extent by the MPs' party affiliations. In comparison to previous findings regarding certain presidential systems, MPs committee assignments or gender were found irrelevant.

**Introduction**

The elected members of legislatures form many types of policy networks. In basic terms, the Members of Parliament (MPs) can be considered as the nodes of a network while relations among the MPs represent the network links. One such policy network is the MPs' private initiatives (bills). When two or more MPs cooperate and co-sign a private initiative, a link is established among the cosponsors of the initiative. When the politicians collaborate an interesting question is who cooperates with whom and why? The main research problem we address here is to try to characterize the structures of private initiative cosponsor networks. In the terminology of Alemán and Calvo (2013) we ask what the determinants of the private initiative policy networks are. Such policy network analyses have previously been carried out regarding certain presidential systems (Alemán and Calvo 2013; Crisp et al. 2004; Zhang et al. 2008; Fowler 2006a; b; Tam Cho and Fowler 2010; Wilson and Young 1997; Koger 2003; Schiller 1995; Kessler and Krehbiel 1996). Here, however, we consider a typical European parliamentary multiparty system. As far as the authors are aware of, no previous studies on the subject exist with respect to parliamentary systems.

European legislatures have been considerably rarely studied in the field of private initiatives when compared with the U.S. that has a long history of scholarly literature analyzing private bills in the U.S. Congress from various points of view. This asymmetry is partly due to differences in approach to the study of MP behavior. In the U.S. the individualist approach highlights the importance of the individual congressman, however, according to the often prevailing party-collectivist approach with respect to (European) parliamentary legislatures the basic unit of research is the party group instead of individual MPs (Esaiasson 2000, 51-52; see also Pajala 2014). Another related explanation refers to the political importance of the private initiatives. The European parliamentary systems mainly operate with government bills thus leaving the private initiatives in a rather marginal role, while in presidential systems the private bills are much more important tools of policy making. The handful of the European studies have almost exclusively focused on analyzing the motivational side of this activity, i.e. why the MPs draft private initiatives in vast amounts knowing that the prospects of getting one passed are almost nil. To address this question on both sides of the Atlantic private bill or initiative drafting has been shown to be strongly tied with the electoral connection theory, according to which the purpose of the drafting is the cultivation of the representatives' personal vote (Mayhew 1974; Koger 2003; Brauninger et al. 2012; Brunner 2013; Solvak 2013). In the European context it has been shown that nuances of the electoral systems seem to provide incentives to draft more/less private initiatives (Bräuninger et al. 2012; Solvak 2011; 2013: Solvak and Pajala 2016).

Instead of the motivational aspects, we follow the small strand of the U.S. spawned research, which has aimed at finding and characterizing private bill cosponsor communities or clusters within the full cosponsor network (Fowler 2006a; b; Zhang et al. 2008; Tam Cho and Fowler 2010; Aleman and Calvo 2013). The main characteristics or determinants in presidential systems have been shown to include MPs' party affiliation, electoral district and committee membership (Alemán and Calvo 2013; Kirkland 2011; Zhang et al. 2008) and to some extent also gender (Clark and Caro 2013) and also Ethnicity (Bratton and Rouse 2011). We shall test whether the above variables (excluding ethnicity) appear as determinants of private initiative policy networks in a parliamentary multiparty system. The analysis is also extended beyond the legislative initiatives by including every type of private initiative, motion or amendment the MPs are allowed to draft in the parliament, in our case the Finnish parliament Eduskunta. The initiatives are analyzed as two sub-categories: one including legislative and the other budgetary initiatives. The legislative initiatives are aimed at statewide legislation and only seldom have specific regional aspects. In contrast, the budget motions are nearly always aimed at specific geographical areas, that is, the MPs' home districts. The budget motions are also a substantial part of the data constituting well over 50 % of all initiatives. As cosponsored private initiatives are introduced in thousands per an electoral term we face a methodological challenge trying to analyze dense policy networks. To overcome the problem we shall adopt a network detection approach introduced by Tumminello et al. (2011). The approach has previously been applied to various types of networks (see Puccio et al. 2016 and references therein). In a nutshell, the approach allows us to see through the noise in a dense network and is able to filter out statistically significant links among MPs in the initiative networks. The procedure results in a number of statistically validated MP clusters or communities that represent the network core. The communities can further be analyzed to reveal which of our attributes appear as the determinants of the communities.

The resulting analyses suggest three main contributions: First, communities found in the legislative initiative networks are first and foremost determined by the party affiliations of the MPs. Second, the electoral district of an MP is the dominant determinant found in the budgetary initiative networks while the MPs' party affiliations explain these communities only to a limited extent. Third, the MPs' committee membership and gender were surprisingly found to be irrelevant attributes.

The paper is organized as follows: after the introduction we take a look at previous wisdom on legislatures, private initiatives and cosponsoring. The next section introduces our network core attributes together with a brief theoretical discussion. Subsequently, we introduce a brief introduction to the private bill system in Eduskunta. The applied data overview together with the network community analyses are in the following section. Finally, a discussion concludes the paper.

**Legislatures, private initiatives and cosponsoring**

We shall hereinafter use the terms (private) initiative, motion or proposal interchangeably.[1] In order to illuminate the differences between presidential and parliamentary systems we shall briefly review key aspect regarding both systems. To start with, the U.S. system does not have formal government bills as legislation must spawn within the Congress. This, by definition, implies the private bills being very important. The congressmen cannot act alone, as they depend on the executive leadership and party leadership (Mattson 1995, 449; Cox and McCubbins 1993). Mattson (1995, 449, footnote 2) further cites Lindblom, who estimated 80 % of the bills enacted into law originating in the executive branch (Lindblom 1968, 88). The executive branch and party leadership act as strong background forces, however the fate of the bills is ultimately in the hands of congressmen. In contrast, in parliamentary systems the legislatures almost exclusively process government bills. For example, Pajala (2012b) concludes 99 % of the government bills being successful and 99 of the private initiatives failing in Finland. Therefore it is no surprise that the private bills or initiatives are sometimes referred to as 'pseudolegislation' (Mattson 1995). Despite the differences between the parliamentary and presidential systems in both representatives draft and cosponsor private initiatives in vast numbers.

A scholarly answer provided to the previous puzzle has its' roots in the representatives' electoral connection and in their personal vote (Mayhew 1974; Cain et al. 1987). This implies the main audience of the initiatives being voters. The private bills are drafted in order to cultivate the representatives' personal vote. Often re-election is considered one of the main motives behind the representatives' activities. Anecdotal evidence supporting this can be found in Schiller (1995) and Koger (2003) including sections based on interview data in the U.S. Congress. There is also evidence about variance in the need for the personal vote depending on the size of the representatives' home constituency, mandate type or variations in electoral systems as certain MPs might have an incentive to draft more private bills than others. Koger (2003) found, to a certain extent, private bill cosponsoring in the U.S. Congress varying with member's electoral circumstances, institutional position and state size. Bräuninger et al. (2012) find support for this claim in the Belgian proportional flexible list system. Similar evidence has been found regarding Finland and Estonia (Solvak 2011, Solvak 2013; Solvak and Pajala 2016). A formal model of private bill introduction is established in Brunner (2012) and tested with data from the French parliament. In the U.S. Congress Schiller (1995) finds a combination of institutional and political forces to constrain Senators in their use of bill sponsorship.

---

[1] Compared with the U.S. Congress the previous terms refer, broadly speaking, to both private *and* public bills. The U.S. public bills aim at establishing federal laws applicable to the general public. These are somewhat equivalent to Finnish legislative and petite motions. The U.S. private bills, in turn, have a narrower scope and address with more specialized matters. In Finland these would come close to the budget motions. Our data include every type of the Finnish private initiatives.

Another scholarly answer to bill drafting and cosponsoring argues the main audience of the private bills to lie within the legislature (Kessler and Krehbiel 1996; Wilson and Young 1997). The legislative connection theory suggests MPs being more interested in the subject matter than credit claiming. Kessler and Krehbiel (1996) set out to test which theory meets the empirical investigation better, the electoral connection or the parliamentary connection and find support for the latter. Also Wilson and Young (1997) argue the private bills being means to signal support, ideological content and expertise. Further, Wilson and Young (1997) show that having many cosponsors does not necessarily help the private bills. Pajala (2012) reports a similar phenomenon regarding Finland.

Studies considering cosponsoring as social networks have been carried out as well. Fowler (2006a) defines a measure of network 'connectedness' in order to find the most influential legislators in the Congress during 1973-2004. Fowler (2006b) is a survey of traditional network centrality measures in addition to the connectedness. Fowler (2006b) discusses connectedness and roll call votes and finds a weak but positive connection between connectedness and vote choice. Further, Alemán et al. (2009) find a strong correlation between roll-call vote and bill cosponsorship patterns in in the U.S. House of Representatives and Argentine Chamber of Deputies. Tam Cho and Fowler (2010) examine the social network structure of Congress and find it exemplifying characteristics of a 'small world' network where actors are densely interconnected with few intermediaries. Zhang et al. (2008), using the concept of modularity, set out to identify the community structure of congressmen, who are connected by private bill cosponsorship. Zhang et al (2008) show the cosponsor patterns to follow the party trench line between the Democrats and the Republicans. Only a few intermediary legislators cosponsor private bills over the party line. The modularity analysis reveals cooperation among congressmen within the same state or neighboring states. This is seen reasonable as many private bills have a pork-barrel nature and are aimed at benefitting the (co)sponsors' home districts. The cosponsor patterns are found to break the party line in one instance where a group of southern Democrats consistently cosponsor legislation with Republicans from the same area. However, the group has significantly diminished in size over time and is nowadays somewhat small (Zhang et al. 2008, 1709). Alemán and Calvo (2013) carry out a cross-national analysis of policy networks in the Latin America (Argentie and Chile). Treating private bill cosponsoring as a social network Alemán and Calvo (2013) result in a probability interpretation, according to which the likelihood of a policy tie is more likely to exist if the MPs share the same party affiliation, are assigned to the same committee or come from the same electoral district. The probability interpretation does not, however, reveal the network core or the network communities within it.[2]

**Theoretical expectations and the network attributes**

The below applied network analysis method does not employ any pre-classification of MPs. The structure of cosponsoring, i.e. the initiatives and the cosigners alone determine the network core communities. On the theoretical side we may ask what kind of communities we should expect to find. The two theories revolving around the private initiatives, Mayhew's (1974) electoral connection and the parliamentary connection theory (see Kessler and Krehbiel 1996; Wilson and Young 1997) provide certain implications. The logic of the personal vote aspect in the former theory suggests the best option to draft the initiatives alone to enjoy the most of the potential benefits. Additional cosponsors from one's own or other party groups lower the potential benefits for a single MP. Our data supports this view as half of the initiatives are not cosponsored. The next best option would be to cooperate with one's party comrades ensuring the potential benefits within the party group. Hence, while only the second best option, the electoral connection theory does not overrule party group cosponsoring. The legislative connection theory suggests the MPs being policy oriented, i.e. in pursue of good public policy (see Fenno 1973; Campbell 1982). The target of the initiatives is within the parliament. This would imply active support seeking among fellow MPs. The cosponsors could thus arise from party group comrades, fellow committee members, representatives from the same sex or form as completely ad hoc support groups. While both theories open up certain cosponsor combinations the aim of the study is to identify which ones are the most stable and recurring.

---

[2] The methodology Alemán and Calvo (2013) apply is essentially to weight bootstrapped exponential random graph model (B-ERGM) estimates by the frequency of observed ties "with few assumptions and without imposing a parametric form that may bias the results".

On the empirical side, we note the well-formulated argument by Alemán and Calvo (2013) that "policy networks should reflect the cohesion of parties, responsiveness to district level principals and jurisdictional expertise". Hence, sharing the same party affiliation, electoral district or committee membership should enhance the probability of policy ties, which reasoning is supported by the results of Alemán and Calvo (2013) and to some degree also by Zhang et al. (2008). As a local starting point, Nousiainen (1961) predicts the existence of four (co)sponsor categories specific to Eduskunta: single MP, party group, multiple party group and electoral district based private initiatives. Later Pajala (2014a), using a simple count analysis, verified the existence of the (co)sponsor categories and found roughly half of the private initiatives to fall into Nousiainen's categories. Whether cooperation based on committee memberships or gender is relevant in Finland, however, remains an open question.

Next we shall outline the network community attributes to be tested. We consider certain specific attributes: MPs' party affiliation, electoral district, committee assignment and gender. We also consider certain general attributes: whether an electoral district is in a metropolitan area (Helsinki and Uusimaa) or in a rural area (all others), the government/opposition status of the party groups as well as its political position (left, center or right wing party).

In Finland as in other parliamentary systems, the party groups in general consist of somewhat likeminded MPs, however, the party groups tend to be rather well disciplined as well. These are prerequisites for maintaining a coalition government, which must enjoy the support of the parliament. Overall the parliamentary systems are very party oriented so it is natural to assume the party affiliation to be one of the key network community determinants. The empirical findings in Pajala (2014a) together with the U.S. findings in e.g. Alemán and Calvo (2013) and Zhang et al. (2008) support this assumption.

While the MPs act as representatives of various parties and ideologies they also act as representatives of their home constituencies and thus various geographical areas of the country. We agree with Aleman and Calvo (2013) stating "legislators elected from the same electoral districts are likely to share preferences for distributive policies that target their constituencies". While legislative initiatives mostly concern with statewide legislation the budget motions are almost exclusively aimed at the MPs' home constituencies (Pajala 2014a). As is obvious this is a fundamental difference. The budget motions seek funding for local infrastructure projects such as roads, bridges or hospitals. The private initiatives tend to be read by ministry officials and in rare occasions the initiatives might later re-appear as parts of government bills (Pekonen, 2011, 179). In this sense, the strongest signal for a constituency project (to be perhaps included in the states' budget later) is a joint budget motion by every MP in the district. In fact, often certain district initiatives are drafted year after year. A Finnish feature is the yearly practice to circulate the drafting responsibility among the parties in the district. Also the findings in Nousiainen (1961) and Pajala (2014a) supporting the electoral district as one of the initiative networks core determinants.

Aleman and Calvo (2013) find shared committee membership increasing the likelihood of cosponsoring bills in the Congress. Somewhat similar result regarding U.S. state level legislature committees is presented by Kirkland (2011). Committee service enhances the MPs' policy expertise and the recurring contacts create opportunities to share information on policy preferences and interests. In Eduskunta, plenary agenda items are first considered by the appropriate committee(s). In individual MP level the committee assignment tend to follow a pecking order, however, are also partly based on self-selection (Forsten 2005). At committee level, however, the compositions of the committees reflect the party groups' seat shares leading to a government-opposition setting inside the committees (Pekonen 2011). Every party group is represented in nearly every committee. This aspect lowers the likelihood of committee membership as a network community determinant as it would require systematic within committee cooperation over the party line. Pajala (2012b) does report a few cases of within committee cooperation among the accepted initiatives, however, these are extremely rare occasions.

Clark and Caro (2013) find evidence for gender based networking in a subset of "women's issues" in multimember districts in the U.S. In a historic perspective, an unofficial women's network has existed in Eduskunta (Pajala 2013; Honka-Hallila 2006). The network, however, refers to the earlier days of the parliament when relatively few women were elected to Eduskunta. Nowadays female MPs constitute roughly 40 % of the representatives and the Nordic countries are known to be the forerunners in equality between the sexes. Pajala (2013) found only a few private bills having exclusively female MP cosponsors during the 1999-2002 parliamentary term. As above, a substantial amount of female/male sponsored initiatives are required should this attribute appear to characterize the network core communities. Moreover, the initiatives must be cosponsored among the same (at least few) MPs.

From time to time the media is eager to describe certain issues as capitol area vs. the rest of the country. The capitol area, located in the south of the country, represents close to 20 % of the population. The geographically small capitol area consists of two electoral districts: Helsinki (the capitol city) and Uusimaa (the surrounding area). This attribute, should it be relevant, would indicate the polarization of the country in a north-south direction. However, rather than other districts of the country it is intuitively easier to assume the capitol area MPs to act in unison.

The general government/opposition attribute has an exploratory nature. It can be triggered in the case of a party affiliation characterized community according to the government/opposition status of the party group(s) in question. The attribute can also be triggered without the party attribute, which suggests parts of government/opposition groups systematically cosponsoring initiatives.

The last general attribute of left-right status operates as the previous one. A party group characterized community can also trigger the left/center/right attribute depending on the political location of the group(s). As above, the attribute can be triggered without the party attribute indicating small parts of certain groups acting in unison.

**Private initiatives in Eduskunta**

Before proceeding further we shall first provide a brief introduction to the Finnish context and its key properties. Finland is a somewhat typical European parliamentary multiparty system. The 200 MPs in Eduskunta are elected every fourth year. The general elections are proportional using open candidate lists. During 1999-2014, the country was divided into 15 electoral districts and the numbers of MPs elected from each one reflect the population sizes. The more densely populated areas in the south of the country constitute relatively small districts while the northern districts are geographically larger and have less MPs. As Table 1 shows the metropolitan area of Helsinki and Uusimaa occupy 55 (27 %) of the seats in the parliament.

**Table 1**. Electoral districts, their abbreviations, and number of elected MPs

| District | Abbrev. | # of elected MPs |
|---|---|---|
| Aland | Alan | 1 |
| Etela-Savo | Esav | 6 |
| Hame | Hame | 14 |
| Helsinki | Hels | 21 |
| Central Finland | Keski | 10 |
| Kymi | Kymi | 12 |
| Lapland | Lapp | 7 |
| Oulu | Oulu | 18 |
| Pirkanmaa | Pirk | 18 |
| North Karelia | Pkar | 6 |
| Pohjois-Savo | Psav | 10 |
| Satakunta | Sata | 9 |
| Uusimaa | Uusi | 34 |
| Vaasa | Vaas | 17 |
| Varsinais-Suomi | Vars | 17 |

Eight political party groups were present in the parliament during 1999-2014. The groups' seat shares together with their governmental status and political positioning are listed in Table 2. A significant change to the otherwise very stable occurred in the 2011 general elections when the populist right-wing Ps group resulted in a landslide victory as can be seen in Table 2. Accordingly, the political landscape now consisted of four large parties (Kesk, Kok, Ps, Sdp) instead of the former three (Kesk, Kok, Sdp). The former three large party setting existed at least from the early 1980s. Governments up to 2010 included a combination of two of the large parties while the third was in the opposition. After the formation of Katainen's "sixpack" government in 2011 the opposition consisted of only two of the large parties (Kesk and Ps).

**Table 2**. Party groups and their political position, seats and government-opposition status

| Party | 1999-2002 | 2003-2006 | 2007-2010 | 2011-2014 |
|---|---|---|---|---|
| Kd (Right-wing) | 10 (4.2%) | 7 (5.3%) | 7 (4.9%) | 6 (4.0%) |
| Kesk (Centre) | 48 (22.4%) | 55 (24.7%) | 51 (23.1%) | 35 (15.8%) |
| Kok (Right-wing) | 46 (21.0%) | 40 (18.6%) | 50 (22.3%) | 44 (20.4%) |
| Ps (Right-wing) | 1 (1.0%) | 3 (1.6%) | 5 (4.1%) | 39 (19.1%) |
| Rkp (Centre) | 12 (5.1%) | 9 (4.6%) | 10 (4.6%) | 10 (4.3%) |
| Sdp (Left-wing) | 51 (22.9%) | 53 (24.5%) | 45 (21.4%) | 42 (19.1%) |
| Vas (Left-wing) | 20 (10.9%) | 19 (9.9%) | 17 (8.8%) | 14 (8.1%) |
| Vihr (Left-wing) | 11 (7.3%) | 14 (8.0%) | 15 (8.5%) | 10 (7.3%) |

Notes: Numbers in parenthesis represent shares of votes in the general elections; Government groups' cells are highlighted gray. Abbreviations: Kd = Christian Democrats, Kesk = Center Party, Kok = National Coalition Party, Ps = True Finns, Rkp = Swedish People's Party, Sdp = Social Democratic Party, Vas = Left Alliance, Vihr = Green League.

The private motion system follows a set of rules, procedures and conventions. First of all, the private initiatives are an individual right. Unlike certain other parliaments there are no party group or committee initiatives. Drawing from Pajala (2011) the right of an MP to introduce motions is included in the Constitution, according to which the MPs have the right to introduce: 1) A legislative motion containing a proposal for the enactment of an Act. 2) A budgetary motion containing a proposal for an appropriation to be included in the state's budget supplementary budget, or for other budgetary decision. 3) A petitionary motion containing a proposal for the government to draft a law or for taking other measures. 4) A topical debate (DEB) be held in a plenary session. Legislative motions can be introduced whenever the parliament is in session and the budgetary motions in connection with the state's annual budget or any supplementary budgets. While the debate proposals are processed by the Speaker's Council all other motions are processed by a standing committee as decided by the plenary. The cosigners must sign the initiatives *before* they are handed to the parliament's central administration. The first signatory of a motion can withdraw the motion without consulting the cosponsors. According to a long standing convention, the cabinet ministers maintain their seats in the parliament, however do not engage in private initiative drafting or cosponsoring.

The private initiative system has undergone only somewhat modest changes over the decades. Among its European peers Eduskunta's private motion system belongs to the more liberal ones. Mattson (1995) surveyed five restrictions with respect to private bill rules and procedures. Regarding Eduskunta these restrictions appear as: First, there are no numerical limits of how many motions the MPs can introduce. Second, the only time limits refer to the budget motions. Third, the only technical requirements refer to legislative initiatives, which have to be written in a form of an act like the government bills. Fourth, there are no limitations on the contents of the motions. Fifth, killing (or burying) a motion in committee after its plenary introduction is the practice used Eduskunta to stop private motions reaching further deliberation. As the parliament first and foremost operates with government bills the private initiatives extremely rarely become laws. Rather many of the legislative initiatives are in fact minor revision suggestions for government bills. As these initiatives have to be processed together with government bills it ensures their consideration in committees and plenary, however, the private bills are passed only in very rare occasions.

**Data and the cosponsor network communities**

Our database consists of every legislative and budget initiative submitted in the Finnish parliament between 1999 and 2014, along with information about the (co)sponsoring MPs including their gender, electoral district, party affiliation and committee membership. The database is split in four subsets, each encompassing a period of a four-year parliamentary term. Each dataset is further divided into legislative and budget initiatives. Table 3 shows the former make up roughly one third of the 1999-2002 dataset, one fifth of the 2003-2006 dataset, one sixth of the 2007-2010 one and one fifth of the 2011-2014 term. While roughly half of the initiatives have only one signature there are also initiatives that connect almost every MP to every other MP having well over 100 cosponsors. Table 3 shows numbers of legislative initiatives to be declining while the number of budget initiatives seems to vary over time. The total number of initiatives is 21 069.

**Table 3**. Summary statistics of legislative and budget initiatives for each parliamentary term

| Term | 1999-2002 | | 2003-2006 | | 2007-2010 | | 2011-2014 | |
|---|---|---|---|---|---|---|---|---|
| Type | legisl. | budget | legisl. | budget | legisl. | budget | legisl. | budget |
| N | 1932 | 3826 | 1208 | 5134 | 956 | 4780 | 587 | 2646 |
| i-range | 1-144 | 1-124 | 1-175 | 1-73 | 1-136 | 1-44 | 1-150 | 1-92 |
| M | 179 | 176 | 186 | 183 | 183 | 175 | 199 | 174 |
| m-range | 2-316 | 2-239 | 5-249 | 2-412 | 2-197 | 2-644 | 2-159 | 2-662 |

Abbreviations: *i-range* is the range of signatures initiatives received (min.-max.); *M* stands for the number of MPs who signed at least 2 initiatives; *m-range* is the range of signatures members affix (min.-max.).

Methodologically, the idea is to derive information about preferential relationships in the parliament by looking at it as a bipartite system consisting of the MPs on one side and the initiatives on the other. In the unfiltered network, a link between two MPs is established when they both sign (cosponsor) the same initiative. Such a network can be very noisy, with many links between any given pair of MPs, so our goal here is to filter out the noise, that is, discriminate "random links" from those carrying real information (intentional collaboration). In order to do so, each link in the network is validated against a null hypothesis, of having drawn that link at random out of the full set. The probability distribution we employ in the null hypothesis is the hypergeometric distribution, which enables us to assign a p-value to each link. With the aim of validating links, we need to set a threshold, and we do so by using criterion at the 1% significance that is corrected with Bonferroni multiple hypothesis test (Tumminello et al. 2011). All the links that fall below the Bonferroni threshold are thus validated.

One matter to take in due consideration is the heterogeneity on the initiatives' side, as an initiative could have been signed by just 2 people, or over a hundred as Table 4 shows. To account for this effect, we decided to divide initiatives in groups, based on the number of people who signed them, before validating links within each group. Groups were set equal to all parliamentary terms, assuming that overall results do not depend strongly on the choice made. We grouped initiatives signed, respectively, by 2, 3, 4, 5, 6, 7, 8, 9, 10 people, by 11 to 13 people, 14-20 people, 21-40, 41-100 and finally those signed by more than a hundred people, for a total of 14 groups. Since a link between the same pair of MPs is likely to be validated in more than one group, we consider the network weighted and assign a weight to each link equal to the number of groups it was validated in. As building a statistically validated network (Bonferroni network hereinafter) between MPs is the goal, all MPs who signed less than two initiatives, as well as initiatives with less than two signatures have no relevance to our analysis and have thus been removed.

**Table 4.** Statistics of Law and Budget Bonferroni Networks for each parliamentary term

| Term | 1999-2002 | | 2003-2006 | | 2007-2010 | | 2011-2014 | |
|---|---|---|---|---|---|---|---|---|
| Network | legisl. | budget | legisl. | budget | legisl. | budget | legisl. | budget |
| $M_B$ | 107 | 157 | 95 | 174 | 62 | 160 | 106 | 121 |
| $L_B$ | 265 | 804 | 687 | 1636 | 146 | 1134 | 1423 | 1663 |
| $f(\%)$ | 1.7% | 5.9% | 4.1% | 14.0% | 0.9% | 15.4% | 7.8% | 23.3% |
| Mean $w$ | 3.6±1.5 | 2.9±1.9 | 3.0±1.4 | 2.2±1.7 | 2.7±1.2 | 2.5±2.0 | 2.5±1.0 | 1.9±1.6 |
| Mean $w_B$ | 1.0±0.2 | 1.5±0.8 | 1.1±0.3 | 1.4±0.7 | 1.0±0.2 | 1.6±0.9 | 1±0 | 1.3±0.7 |
| $w_B$-range | 1-3 | 1-7 | 1-3 | 1-6 | 1-2 | 1-9 | 1 | 1-8 |
| $Q$ | 0.75 | 0.59 | 0.55 | 0.53 | 0.64 | 0.68 | 0.55 | 0.49 |

Abbreviations: $M_B$ is the number of members involved in the Bonferroni Network; $L_B$ is the number of validated links; $f(\%)$ stands for the fraction (in percentage) of validated links out of the original number of links; Mean $w$ is the average weight of links in the original network; Mean $w_B$ is the average weight of validated links; $w_B$-range is the range of link weights in the Bonferroni Network (min.-max.); $Q$ is the modularity of the partition found, the closer it is to 1, the better the partition is.

Once the cosponsor network is built, our main interest lies in finding out if and how it is internally organized in communities and ultimately, which attributes characterize each community and to what extent. To this end, the first step is choosing a suitable community detection algorithm. In our case, the choice fell on the software Radatools[3], which employs a combination of different algorithms, allows for weighted networks, multiple repetitions of each algorithm and produces high modularity for the partitions found[4].

After a stable partitioning of the system is obtained, we look for each community's characterizing attributes by validating, community-wise, all MPs attributes available in our datasets. This is accomplished in the same fashion as what we have done when validating links in the network.

---

[3] Radatools v3.2, Copyright © 2011 by S. Gomez, A. Fernandez, J. Borge-Holthoefer and A. Arenas. All rights reserved.

[4] The best results were obtained with 200 repetitions of: e (extremal optimization), b (fine-tuning by bootstrapping based on tabu search), r (fine-tuning by reposition), f (fast algorithm), b, r.

Among the legislative initiative networks, the network detection procedure identified 6-10 communities depending on the parliamentary term as Table 5 shows. The most striking result is the communities being characterized only by two of our specific attributes: the MPs' party affiliation and on two cases also by the electoral district. Committee membership or gender do not characterize any community and have been omitted in Table 5. Among the general attributes the political left/center/right positioning of party groups and the coalitional status are triggered. The (metropolitan/rural) area attribute does not characterize any community and has been omitted in Table 5. Two small communities have no characterization during the 1999-2002 period. As an example, consider the largest 30 member community in the top left quadrant in Table 5 representing the 1999-2002 term. It is characterized, or rather over-represented, by MPs from the Kesk party. Two general attributes are also triggered: the coalitional status as Kesk was an opposition party and the political positioning, which in this case is the center. The next largest 16 member community is the only cluster in Table 5 for which the sole characterizing variable is the political party positioning. The community is over-represented by MPs from the leftist parties, however, over-representation by any specific party group is not identified. Worth noting are communities in the low right quadrant of Table 5 representing the 2011-2014 parliamentary term. The largest 40 member community is homogeneously represented (rather than over-represented) by MPs from the Ps party group. The next largest community to the right is homogeneously represented by the MPs from the Kesk party group, which was the one of the two opposition groups during this term. The result indicates hard opposition competition rather than cooperation. As an overall result, the numbers of detected communities seem to be declining over time. Other details of the legislative communities can be found in Table 5.

**Table 5**. Legislative network communities and their characterizing attributes during 1999-2014

| 1999-2002 – Law Bonferroni Network | | | | | 2003-2006 – Law Bonferroni Network | | | | |
|---|---|---|---|---|---|---|---|---|---|
| $N_C$ | Party | District | Coal. | PPos. | $N_C$ | Party | District | Coal. | PPos. |
| 30 | Kesk | | Opp. | Centre | 40 | Kok | | Opp. | Right |
| 16 | | | | Left | 19 | Vas | | Opp. | Left |
| 13 | Sdp | | | Left | 14 | Vihr | | Opp. | Left |
| 12 | Kok | | | Right | 6 | Kd | | | Right |
| 12 | Kesk | | Opp. | Centre | 4 | Rkp | | | |
| 10 | Kd | | Opp. | Right | 3 | H.Kesk | | | |
| 7 | Vihr | | | Left | 3 | Ps | | | |
| 3 | | | | | 2 | H.Kesk | | | |
| 2 | H.Kesk | H.Vaas | | | 2 | H.Sdp | | | |
| 2 | | | | | 2 | H.Kesk | | | |
| 2007-2010 – Law Bonferroni Network | | | | | 2011-2014 – Law Bonferroni Network | | | | |
| $N_C$ | Party | District | Coal. | PPos. | $N_C$ | Party | District | Coal. | PPos. |
| 17 | Vas | | Opp. | Left | 40 | H.Ps | | H.Opp. | H.Right |
| 12 | Sdp | | Opp. | Left | 36 | H.Kesk | | H.Opp. | H.Centre |
| 10 | Sdp | | Opp. | Left | 21 | Sdp | | Gov. | Left |
| 7 | Kd | | Opp. | Right | 5 | H.Kok | | H.Gov | H.Right |
| 6 | Sdp | | | | 2 | H.Kok | | H.Gov | H.Right |
| 3 | H.Sdp | | | | 2 | H.Vas | | H.Gov | H.Left |
| 3 | H.Kok | | | | | | | | |
| 2 | H.Kok | H.Pirk | | | | | | | |
| 2 | H.Sdp | | | | | | | | |

Abbreviations: $N_C$ = size of community; *Party* = Party affiliation, *District* = electoral district, *Coal.* = government /opposition, *PPos.* = Political position of party group, H. = homogeneously represented.

Turning into the budget networks the detection procedure identified 5-12 communities depending on the parliamentary term as Table 6 shows. Also these network communities are characterized only by the same specific attributes as the legislative ones. The dominating attribute is clearly the electoral district, however party affiliation seems to play a role as well. It appears that budgetary matters break the party trench lines. Only during the most recent term of 2011-2014 on the low right quadrant of Table 6 the party affiliation is dominating. The largest community is over-represented by MPs from the Kesk group while the next largest cluster is homogeneously represented by the members of the Ps group. Among the general attributes the metropolitan/rural attribute is characterizing five communities. As another example in the top left quadrant representing the 1999-2002 term the largest 25 member community is over-represented by MPs from the Uusimaa district. Two of the general attributes are triggered as well as the community is over-represented by MPs from government groups from the metropolitan area.

With respect to both legislative and budget initiative networks in Tables 5 and 6 the community structure seem somewhat different during the most recent term compared with previous terms as the numbers of detected communities appear to be at minimum. This phenomenon is largely system specific having its' roots in the 2011 general elections. The sudden rise of the Ps party changed the structure of the parliament from three large parties into four. The resulting government left only two of the remaining large parties (Kesk and Ps) in opposition. While the opposition parties were competing with the government they were also hard rivals. Such a drastic change in the structure of the parliament probably influenced the inner workings of the parties. Not only did the parties close ranks regarding legislation but also regarding budgetary policy making. It appears significant changes in the power balance among the parties can even affect the structure of private initiatives.

**Table 6**. Budget network communities and their characterizing attributes during 1999-2014

| \multicolumn{5}{c|}{1999-2002 – Budget Bonferroni Network} | \multicolumn{5}{c|}{2003-2006 – Budget Bonferroni Network} |
|---|---|---|---|---|---|---|---|---|---|
| $N_C$ | Party | District | Area | Coal. | PPos. | $N_C$ | Party | District | Area | Coal. | PPos. |
| 25 |  | Uusi | Met. | Gov. |  | 50 | Kok | Vars, Esav |  | Opp. | Right |
| 24 | Kesk | Esav, Psav | Rur. | Opp. | Centre | 34 | Vihr | Uusi | Met. |  | Left |
| 18 |  | Vaas |  |  |  | 26 | Vas | Lapp, Psav |  |  | Left |
| 12 |  | Vars |  |  |  | 18 |  | Vaas |  |  |  |
| 12 | Vihr |  |  |  |  | 16 |  | Hame, Keski |  |  |  |
| 12 |  | Pirk |  |  |  | 14 |  | Pirk |  |  |  |
| 10 |  | Sata |  |  |  | 8 |  | Kymi |  |  |  |
| 10 | Kd |  |  | Opp. | Right | 6 |  | Pkar |  |  |  |
| 9 |  | Kymi |  |  |  | 2 | H.Kesk | H.Oulu |  |  |  |
| 9 |  | Keski |  |  |  |  |  |  |  |  |  |
| 9 |  | Pkar |  |  |  |  |  |  |  |  |  |
| 7 |  | Hame |  |  |  |  |  |  |  |  |  |
| \multicolumn{5}{c|}{2007-2010 – Budget Bonferroni Network} | \multicolumn{5}{c|}{2011-2014 – Budget Bonferroni Network} |
| $N_C$ | Party | District | Area | Coal. | PPos. | $N_C$ | Party | District | Area | Coal. | PPos. |
| 38 | Sdp | Keski, Kymi, Psav | Rur. |  |  | 42 | Kesk |  |  | Opp. | Centre |
| 30 |  | Uusi | Met. |  |  | 36 | H.Ps |  |  | H.Opp. | H.Right |
| 20 | Vas |  |  | Opp. | Left | 16 | Kd | Pirk |  | Gov. |  |
| 19 | Rkp | Vaas |  |  | Centre | 15 | Rkp | Vaas |  |  |  |
| 18 |  | Pirk |  |  |  | 12 |  | Hame |  |  |  |
| 14 |  | Vars |  |  |  |  |  |  |  |  |  |
| 12 |  | Hame |  |  |  |  |  |  |  |  |  |
| 6 |  | Esav |  |  |  |  |  |  |  |  |  |
| 3 |  | Lapp |  |  |  |  |  |  |  |  |  |

Abbreviations: $N_C$ = size of community; *Party* = Party affiliation, *District* = electoral district, *Area* = metropolitan vs. rural, *Coal.* = government /opposition, *PPos.* = Political position of party group, H. = homogeneously represented.

**Discussion**

This article complements previous findings on the private initiatives in European parliamentary legislatures as well as certain previous results regarding Finland. Instead of analyzing incentives of why the MPs draft private initiatives our results shed light on the patterns of cosponsoring the initiatives. Among the legislative and budgetary networks, which are analyzed separately, the core communities of the former are mostly characterized by the MPs' party affiliations. To a lesser extent, the communities are also characterized by the MPs' government/opposition status and the political positioning of the MPs' party groups. The budgetary networks are mostly characterized by geographical cooperation i.e. the MPs' home constituencies and to a lesser extent by the party affiliation. The latter, however, seems to be on the rise over time. In certain cases the MPs' status as capitol area representative or as a representative from the rest of the country, governmental status and political positioning appeared relevant community attributes as well. With respect to legislative and budgetary networks MPs' gender and committee assignments were found irrelevant. Outside Europe, the article also complements previous findings regarding certain presidential systems. The MPs' party affiliation and geographical attributes appear to be relevant in both systems, however departing from the presidential legislatures the committee assignments seem to be irrelevant –at least in the case of the Finnish parliament.

Our results should be generalized to other parliamentary multiparty systems with caution as country specific features might considerably affect the results. For example, in contrast to certain other European legislatures committee initiatives do not exist in Finland. Moreover, the committees resemble mini-plenaries having an internal government-opposition setting, which most likely render the committee attribute irrelevant in Finland. The German Bundestag provides another example with somewhat different initiative features allowing party initiatives rather than initiatives drafted by single MPs.

In reflection to the introduction, we make a remark to the discussion regarding the party-collectivist and the currently applied individualistic approaches to the study of MP behavior. While the former is very often appealing and prevailing in European, or European influenced legislatures, the individualistic approach is capable of providing interesting insights to the study of parliamentary decision making as well. While our results show MPs party affiliation to be a dominant feature in the overall MP collaboration, the results also show the existence of other type of systematic cooperation.

Finally, data collecting has become significantly easier during the last few years as many European and other parliaments have considerably developed their online data availability. In fact, very recently Briatte (2016) has provided private initiative data together with open source software for collecting the data for twenty parliaments. The above applied network detection methodology is well-suited for a comparative cross country analysis, which unfortunately is beyond the scope of this study.